\documentclass[prd,a4paper,nofootinbib]{revtex4}
\usepackage{mathrsfs}
\usepackage{graphicx}
\usepackage{dcolumn}
\usepackage{bm}
\usepackage{epsfig}
\usepackage{slashed}
\newcommand{\be}{\begin{equation}}
\newcommand{\ee}{\end{equation}}
\newcommand{\bea}{\begin{eqnarray}}
\newcommand{\eea}{\end{eqnarray}}
\newcommand{\bean}{\begin{eqnarray*}}
\newcommand{\eean}{\end{eqnarray*}}

\newcommand{\gapproxeq}{\lower
.7ex\hbox{$\;\stackrel{\textstyle >}{\sim}\;$}}
\newcommand{\lapproxeq}{\lower
.7ex\hbox{$\;\stackrel{\textstyle <}{\sim}\;$}}

\begin{document}

\bibliographystyle{unsrt}

\title{ The lineshape of $\psi(3770)$ and low-lying vector charmonium resonance parameters in  $e^+ e^-\to D\bar D$}

\author{Yuan-Jiang Zhang$^{1}$\footnote{E-mail: yjzhang@ihep.ac.cn},
and Qiang Zhao$^{1,2}$\footnote{E-mail: zhaoq@ihep.ac.cn}}

\affiliation{1) Institute of High Energy Physics, Chinese Academy of
Sciences, Beijing 100049, People¡¯s Republic of China}

\affiliation{2) Theoretical Physics Center for Science Facilities,
CAS, Beijing 100049, People¡¯s Republic of China}

\date{\today}

\begin{abstract}
We investigate the $D\bar{D}$ production in $e^+e^-$ annihilations
near threshold in an effective Lagrangian approach. This shows that
the lineshape of the cross section near threshold is sensitive to
the contributions from $\psi^\prime$, though it is below the
$D\bar{D}$ threshold. The recent experimental data from the BES and
Belle collaborations allow us to determine the $\psi^\prime
D\bar{D}$ coupling constant which appears to be consistent with
other theoretical studies. As a consequence, the
$\psi^\prime$-$\psi(3770)$ mixing parameter can be extracted around
the $\psi(3770)$ mass region. Resonance parameters for $\psi(3770)$,
$X(3900)$, $\psi(4040)$, and $\psi(4160)$ are also investigated. The
$X(3900)$ appears as an enhancement at around 3.9 GeV in the Belle
data. In addition to treating it as a resonance, we also study the
mechanism that the enhancement is produced by the $D\bar{D^*}+c.c.$
open channel effects. Our result shows that such a possibility
cannot be eliminated.

\end{abstract}

\maketitle


\vspace{1cm}

\section{Introduction}

As the first charmonium state above the $D\bar{D}$ threshold, the
production of $\psi(3770)$ in $e^+ e^-\to D\bar{D}$ serves as a
peculiar probe for exploring the QCD dynamics in the interplay of
the perturbative and the non-perturbative regime. During the past
few years experimental measurements were performed in the
$\psi(3770)$ region and some interesting observations were exposed.
First, BES-II reported a branching ratio for $\psi(3770)\to$
non-$D\bar{D}$ up to 15\%~\cite{Ablikim:2007zzb,Ablikim:2008zzc},
while CLEO-c found a much smaller non-$D\bar{D}$ branching
ratio~\cite{Besson:2005hm}. Note that so far only one exclusive
channel $\psi(3770)\to \phi\eta$ has been observed in $\psi(3770)$
non-charmoniun strong decays. One major concern is how the light
hadrons are produced in $c\bar{c}$ annihilation. In other words,
whether this process is dominated by pQCD or there are signs for
non-pQCD contributions are questions for both experiment and theory.

Further interests in this issue were raised by a recent
non-relativistic QCD (NRQCD) calculation to the next-to-leading
order (NLO) for the $\psi(3770)$ non-$D\bar{D}$
decays~\cite{He:2008xb}, where the authors found significant QCD
corrections from NLO. Implications of such a result would be that
non-pQCD mechanisms may start to play a role. Quantitative studies
of non-pQCD mechanisms are presented in
Ref.~\cite{Zhang:2009kr,Liu:2009dr}, where the authors show that the
long-distance interactions due to intermediate $D$ meson loops are
essential for understanding the non-$D\bar{D}$ decays of the
$\psi(3770)$. This also suggests that a dynamic understanding of the
correlation between the $\psi(3770)$ non-$D\bar{D}$ decay and the
so-called Okubo-Zweig-Iizuka rule evading~\cite{OZI} mechanism is
required.

$\psi(3770)$ production in $e^+ e^-$ annihilation is also useful for
further studying the properties of the $\psi(3770)$. As shown by BES
measurement~\cite{Ablikim:2008zz,Ablikim:2008zza}, an obvious
deviation of the resonance excitations from Breit-Wigner is observed
in $e^+ e^- \to D\bar{D}$. This raises the question about the
sources of causing such a deviation, and the role played by
background processes \cite{Yang:2006nh}. One possibility is that
some new structures in the energy region between 2.70 and 3.87 GeV
may cause such a lineshape anomaly~\cite{Ablikim:2008zza}. Above the
$\psi(3770)$ mass, the data from Belle Collaboration suggest an
enhancement around 3.9 GeV~\cite{Pakhlova:2008zza}, which could be a
signal for resonance. Note that 3.9 GeV is at the open channel
threshold for $D\bar{D^*}+c.c.$ Thus, it would be interesting to
investigate the $D\bar{D^*}+c.c.$ open channel effects and compare
them with the Breit-Wigner solution in the numerical fits. By
clarifying these issues in both experiment and theory we expect that
the $\psi(3770)$ resonance parameters can be determined.
Furthermore, the recent controversial results from
BES-II~\cite{Ablikim:2007zzb,Ablikim:2008zzc} and
CLEO-c~\cite{Besson:2005hm} on the $\psi(3770)$ non-$D\bar{D}$
decays can be disentangled~\cite{Zhang:2009kr,Liu:2009dr}.

In this work we will study the lineshape of the $e^+ e^-\to D\bar D$
cross section in an effective Lagrangian approach. We will show that
the cross section experiences important interferences from the
$\psi^\prime$ which will account for the $\psi(3770)$ lineshape
deviation from the Breit-Wigner form. A recent analysis of this
issue was also done by Ref.~\cite{Li:2009pw}.  In our work, we will
show that the near threshold cross sections not only provide
evidence for the $\psi^\prime$-$\psi(3770)$ interferences, but also
provide a peculiar constraint on the dynamics for $\psi^\prime$
couplings to $D\bar{D}$. By determining the $g_{\psi^\prime
D\bar{D}}$ coupling, we can then examine the $\psi(2S)$-$\psi(1D)$
mixing for the $\psi^\prime$ and $\psi(3770)$, and extract the
mixing angle at the mass of the
$\psi(3770)$~\cite{Rosner:2004wy,Zhang:2009kr}. We will study the
energy-dependence of the mixing parameter. It is essential to keep
unitarity, and further dynamical information about the state
evolutions could be gained.

We will also investigate the bump around 3.9 GeV observed by
Belle~\cite{Pakhlova:2008zza}. Given the success of the potential
quark model (see Ref.~\cite{Eichten:2007qx} for a recent review), a
vector charmonium with $J^{PC}=1^{--}$ at 3.9 GeV will cause great
concern about the non-relativistic $c\bar{c}$ phenomenology. We will
show that this enhancement may be caused by the $D\bar{D^*}+c.c.$
open channel effects.

This work is organized as follows: In Sec. II, we formulate the
transition amplitudes for $e^+ e^-\to D\bar{D}$ from different
sources, which include charmonium resonance excitations,
electromagnetic (EM) background from the vector meson dominance
(VMD) model and open $D\bar{D^*}+c.c.$ effects via intermediate
meson loops. In Sec. III, we present our numerical results along
with experimental observables. A summary will be given in Sec. IV.

\section{The model}

The ingredients considered in this work include charmonium resonance
excitations and EM background (Fig.~\ref{fig-1}), and open-charm
effects via intermediate meson loops (Fig.~\ref{psi_dd}). The
charmonium resonance excitations are constructed in the VMD
model~\cite{Bauer:1975bw,Bauer:1977iq,Li:2007au}, where the EM field
will be decomposed into vector meson fields with both isospin 0 and
1 components. By taking away explicitly the nearby resonance
contributions from e.g. $\psi(3770)$ and $\psi^\prime$, the EM
background contributions arising from the continuum part can thus be
parameterized by an effective coupling and minimized in the
numerical fit.

We also examine the $D\bar{D}$ final state interactions in $e^+
e^-\to \gamma^*\to \psi^\prime\to (D\bar{D}, \ D\bar{D^*}+c.c., \
D^*\bar{D^*})\to D\bar{D}$ due to intermediate $D\bar{D}$,
$D\bar{D^*}+c.c.$, and $D^*\bar{D^*}$ meson loops as shown in
Fig.~\ref{psi_dd}. These contributions do not double-count the
$\psi^\prime$ excitation in Fig.~\ref{fig-1} since a Breit-Wigner is
explicitly introduced there. The effects of such a loop contribution
may give rise to both energy dependence of the $\psi^\prime$ total
width and a relative phase to the QED diagram. However, we note in
advance that these contributions are negligibly small. It also
should be clarified that there is no need to consider the
intermediate meson loop contributions for $\psi(3770)$ since the
effective coupling $g_{\psi(3770) D\bar{D}}$ is extracted from
experimental data and should have included the loop effects.

To study the $D\bar{D^*}+c.c.$ open channel effects, we apply the
experimental data for $e^+ e^- \to D\bar{D^*}+c.c.$ from
Belle~\cite{Abe:2006fj}. This is equivalent to isolating out the
$D\bar{D^*}+c.c.$ open channel contributions from all possible
sources, and can be compared with the Breit-Wigner fit.

In order to evaluate the diagrams of Figs.~\ref{fig-1} and
\ref{psi_dd}, the following effective Lagrangians are needed:
 \bea \nonumber \mathcal{L}_{\mathcal V D\bar
D} &=& g_{ \mathcal V  D\bar D}\{D
\partial_\mu {\bar D}-\partial_\mu D\bar{D} \}{\mathcal V ^\mu},\\
\nonumber \mathcal{L}_{\mathcal V D {\bar D}^\ast} &=& -i
g_{\mathcal V D {\bar D}^\ast} \epsilon_{\alpha\beta\mu\nu}
\partial^\alpha \mathcal{V^\beta} \partial^\mu \bar {D^\ast}^\nu D + h.c. ,\\
\nonumber \mathcal{L}_{\mathcal P D^\ast {\bar D}^\ast} &=& -i
g_{\mathcal P D^\ast {\bar D}^\ast} \epsilon_{\alpha\beta\mu\nu}
\partial^\alpha \mathcal{{D^\ast}^\beta} \partial^\mu \bar {D^\ast}^\nu \mathcal P + h.c. ,\\
\nonumber
 \mathcal{L}_{\mathcal P \bar D D^\ast} &=& g_{D^\ast \mathcal P
 \bar D}\{\bar D
\partial_\mu {\mathcal P}-\partial_\mu {\bar D} \mathcal
P\}\mathcal{D^\ast}^\mu,\\
\mathcal{L}_{\mathcal V D^\ast {\bar D}^\ast} &=& g_{\mathcal V
D^\ast {\bar D}^\ast} \mathcal V ^\mu [({D^\ast}^\nu
\overrightarrow{\partial_\mu} {{\bar D}^\ast}_\nu - {D^\ast}^\nu
\overleftarrow{\partial_\mu} {{\bar D}^\ast}_\nu) - {D^\ast}^\nu
\overrightarrow{\partial_\nu} {{\bar D}^\ast}_\mu + {D^\ast}^\mu
\overleftarrow{\partial_\nu} {{\bar D}^\ast}_\nu]
 \label{lagrangian} \ , \eea
where $ \mathcal P $ and $ \mathcal V$ are the pseudoscalar and
vector mesons, and $\epsilon_{\alpha\beta\mu\nu}$ is the
antisymmetric tensor.

\subsection{Charmonium excitations in VMD and EM background contribution}

According to VMD~\cite{Bauer:1975bw,Bauer:1977iq}, the EM current
can be decomposed into two parts. One is the hadronic part
containing a complete sum over all isospin-0 and 1 vector meson
fields, while the other is the so-called ``bare photon" field. An
empirical role played by the EM "bare photon" field is to assure the
proper normalization of the physical photon field. Its contribution
is generally small. Therefore, this part is minimized in the VMD
model.

\begin{figure}[ht]
\scalebox{0.7}{\includegraphics{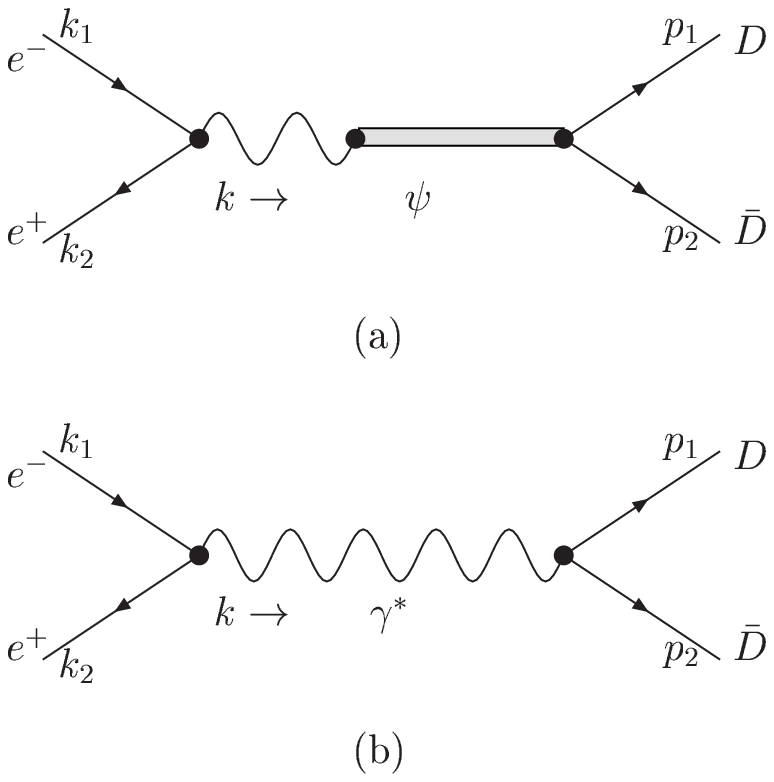}}
\caption{Feynman diagrams based on the VMD model for $e^+e^-\to
D\bar{D}$ near threshold. Diagram (a) is for hadronic vector meson
contributions, while (b) represents the EM background.}
\label{fig-1}
\end{figure}

In reality, it is not possible to include all hadronic vector meson
amplitudes. A commonly adopted method is to include the vector meson
resonances in the vicinity of the considered kinematics, and then
treat the unknown part as an EM background which now includes the
off-shell contributions from those faraway resonances and the ``bare
photon" amplitude. It would also be our strategy here to apply the
VMD model to $e^+e^-\to D\bar{D}$ near threshold.
Figure.~\ref{fig-1}(a) and (b) demonstrate the hadronic
contributions and the EM background, respectively.

Setting $m_e \simeq 0$, the coupling of vector mesons with the
virtual photon can be extracted from experimental data by the VMD
model ~\cite{Bauer:1975bw,Li:2007au}:
 \bea
\frac{e}{f_\psi}=\left[\frac{3\Gamma_{\psi\to e^+ e^-}}{2\alpha_e
|\vec{p}_e|}\right]^{1/2} , \label{eq2} \eea  where $|\vec{p}_e|$ is
the electron three-momentum in the vector meson rest frame,
$\Gamma_{\psi \to e^+e^-}$ is the partial decay width, and
$\alpha_e= 1/137$ is the fine-structure constant.

The amplitude of Fig.~\ref{fig-1}(a) due to $\psi$ resonance
excitations can be written as:
\begin{eqnarray}
T_a &=& e^2 \bar{v}(k_2)\gamma_\mu u(k_1) \frac{1}{s}
\frac{m_\psi^2}{f_\psi}
\frac{1}{s-m_{\psi}^2+im_{\psi}\Gamma_{\psi}} g_{\psi
D\bar{D}}(p_1-p_2)^\mu
 ,\label{Ta}
\end{eqnarray}
where $\Gamma_{\psi}$ is the total width of the charmonium
resonance, and $u(k_1)$ and $v(k_2)$ are the Dirac spinors for the
electron and positron, respectively. The four-vectors $p_1$ and
$p_2$ are the momenta for the final state $D$ and $\bar D$ meson.

We parameterize the EM background contributions to $e^+e^- \to D\bar
D$ as follows:
\begin{equation}
T_b = e^2  \bar{v}(k_2)\gamma_\mu u(k_1) \frac{1}{s}
g_{c}(s)(p_1-p_2)^\mu ,\label{Tb}
\end{equation}
where $g_c$ is an effective coupling of the EM background
contribution to $D\bar{D}$. As pointed earlier, this amplitude
contains the ``bare photon" contribution and contributions from
other vectors which are not explicitly included. In this sense, the
value for $g_c$ will depend on how many vector charmonium states are
included in the fitting. The inclusion of more higher states near
threshold will minimize this amplitude.

Combining Eqs.(~\ref{Ta}) and (~\ref{Tb}), one obtains the total
amplitude for $e^+e^- \to D\bar D$
\begin{eqnarray}
T = \frac{e^2\bar{v}(k_2)\gamma_\mu u(k_1)(p_1-p_2)^\mu}{s} [
g_c(s)+ \frac{m_{\psi^\prime}^2}{f_{\psi^\prime}}
\frac{g_{\psi^\prime D\bar{D}}}{s-m_{\psi^\prime}^2
+im_{\psi^\prime}\Gamma_{\psi^\prime}}  +
\sum_{\psi_i}\frac{m_{\psi_i}^2}{f_{\psi_i}} \frac{g_{\psi_i
D\bar{D}}}{s-m_{\psi_i}^2 +im_{\psi_i}\Gamma_{\psi_i}}e^{i\phi_i}],
\end{eqnarray}
where the $\psi^\prime$ amplitude is explicitly included, and a
phase factor $e^{i\phi_i}$ is added to other charmonium resonance
amplitudes above the $D\bar{D}$ threshold. Then, the cross section
for $e^+e^- \to D\bar D$ can be written as
\begin{eqnarray}
\sigma (e^+e^-\to D \bar{D}) =\frac{8 \pi
\alpha_e^2}{3}\frac{|\vec{p}|^3}{s^{5/2}}|g_c(s)+
\frac{m_{\psi^\prime}^2}{f_{\psi^\prime}} \frac{g_{\psi^\prime
D\bar{D}}}{s-m_{\psi^\prime}^2
+im_{\psi^\prime}\Gamma_{\psi^\prime}}  +
\sum_{\psi_i}\frac{m_{\psi_i}^2}{f_{\psi_i}} \frac{g_{\psi_i
D\bar{D}}}{s-m_{\psi_i}^2
+im_{\psi_i}\Gamma_{\psi_i}}e^{i\phi_i}|^2, \label{cross:section}
\end{eqnarray}
 where $|\vec{p}|$ is the $D$-meson three-vector momentum in the overall c.m. frame.

\subsection{Intermediate meson loop contributions and $D\bar{D^*}+c.c.$ open channel effects}

\begin{figure}[ht]
\scalebox{0.5}{\includegraphics{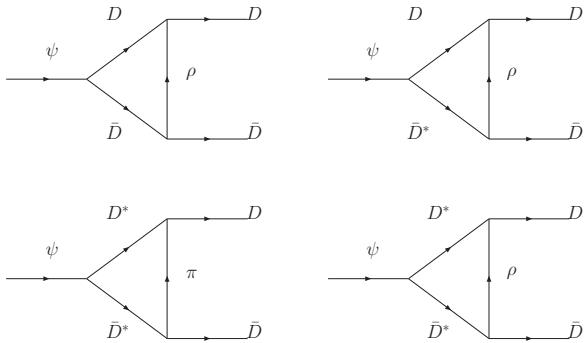}}
\caption{Feynman diagrams for intermediate meson loops as
corrections for the $D\bar{D}$ coupling to a charmonium state. }
\label{psi_dd}
\end{figure}

The intermediate meson loops via the $D\bar{D}$, $D\bar{D^*}+c.c.$
and $D^*\bar{D^*}$ rescatterings in Fig.~\ref{psi_dd} can be
evaluated as follows~\cite{Zhang:2008ab,Zhang:2009kr}:
 \be T_L= -i e^2
\bar{v}(k_2)\gamma_\mu u(k_1) \frac{1}{s} \frac{m_\psi^2}{f_\psi}
\frac{1}{s-m_{\psi}^2+im_{\psi}\Gamma_{\psi}} \int \frac
{d^4p_2}{(2\pi)^4}\sum_{\text{polarization}}
 \frac {T_1T_2T_3}{a_1a_2a_3}{\cal F}(p_2^2) \ .
 \ee
Taking $D \bar D (\rho)$ as a example, the vertex functions for the
$D\bar{D}(\rho)$ loop are
 \be \left\{\begin{array}{ccl}
 T_1 &\equiv& ig_1(l_1-l_3)\cdot \varepsilon_\psi \\
 \nonumber
 T_2&\equiv& ig_2 (l_1 + p_1)\cdot \varepsilon_\rho \\
 T_3&\equiv& ig_3 (l_3 + p_2)\cdot \varepsilon_\rho^*\end{array}\right.
 \ee
where $g_1$, $g_2$, and $g_3$ are the coupling constants at the
meson interaction vertices (see Fig. \ref{psi_dd}). The four-vector
momentum, $l_1$, $l_2$, and $l_3$ are for the intermediate mesons,
respectively, while $a_1=l_1^2-m_1^2, a_2=l_2^2-m_2^2$, and
$a_3=l_3^2-m_3^2$ are the denominators of the propagators of
intermediate mesons.

Divergences are inevitable in the loop integrals. Since the
effective Lagrangian approach is not a renormalizable theory, a
common way to kill the divergences is to introduce a form factor as
a cut-off for those unphysical contributions from the ultraviolet
regime. Such a prescription will also compensate the internal
particle off-shell effects. The form factor ${\cal F}(p^2)$ is
usually parameterized as
 \be
 {\cal F}(p^2) = \left(\frac {\Lambda^2 - m_{ex}^2}{\Lambda^2 -
 p^2}\right)^n, \label{form-factor}
\ee
 where $n=0, 1, 2$ correspond to different treatments of the loop
integrals. If the on-shell approximation is
applicable~\cite{Locher:1993cc,Li:1996yn,Li:1996cj}, the loop
integrals will not suffer from the divergences, and the form factor
will take care of the off-shell effects. In the present work, we
consider the dipole form factor, i.e. $n=2$, in the full loop
integral. The cut-off energy $\Lambda$ is usually parameterized as
 \be \Lambda = m_{ex} + \alpha \Lambda_{QCD}, \ee
where $\Lambda_{QCD}=220$ MeV, $\alpha$ is a tunable parameter, and
$m_{ex}$ is the mass of the exchanged meson. There are different
expressions for the form factor adopted in the literature. However,
note that the cut-off is always tunable. We would generally need
experimental data to determine a proper form factor parameter.

The charmed meson couplings to the light mesons are obtained in the
chiral and heavy quark limits~\cite{Cheng:2004ru},
 \bea g_{D^\ast D
\pi} &=& \nonumber \frac{2}{ f_\pi }\, g \, \sqrt{m_D m_{D^\ast}},
\,\,\,\,\,\,\,\, g_{D^\ast D^\ast \pi} =
\frac{g_{D^\ast D \pi}}{\tilde{M_D}} ,\\
g_{D^\ast D \rho} &=& \sqrt{2} \lambda g_\rho ,\,\,\,\,\,\,\,\,g_{D
D \rho} = {g_{D^\ast D \rho}} \tilde{M_D} ,\eea
 with $f_\pi$ = 132
MeV,\, and $\tilde{M_D} \equiv \sqrt{m_D m_{D^\ast}}$ is a mass
scale. The parameter $g_\rho$ respects the relation $g_\rho =
{m_\rho / f_\pi}$~\cite{Casalbuoni}. We take $\lambda = 0.56\,
\text{GeV}^{-1} $ and  $g = 0.59$ ~\cite{Yan92}.

\begin{figure}[ht]
\scalebox{0.6}{\includegraphics{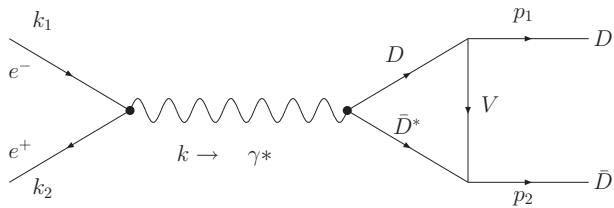}}
\caption{ Feynman diagram for the $D\bar{D^*}+c.c.$ open channel
effects.} \label{open-channel}
\end{figure}

The $D\bar{D^*}+c.c.$ open channel effects (see
Fig.~\ref{open-channel}) can be calculated in a similar way. The
experimental data from Belle for $e^+ e^- \to
D\bar{D^*}+c.c.$~\cite{Abe:2006fj} will allow us to extract the
$\gamma^* D\bar D^* $ coupling $g_{\gamma^* D
\bar{D^*}}(s)$~\cite{Zhang:2008ab}:
\begin{eqnarray} &&\sigma (e^+e^-\to D \bar{D}^* + c.c.) =\frac{4
\pi}{3}\frac{|\vec{p}|^3}{s^{3/2}}\alpha_e^2|g_{\gamma^* D
\bar{D^*}}(s)|^2, \label{crosssection1}
\end{eqnarray}
 where $|\vec{p}|$ is the $D$-meson three-vector momentum in the overall c.m. frame.

As briefly mentioned earlier, explicitly adopting the experimental
data for $e^+ e^- \to D\bar{D^*}+c.c.$ means that all the resonances
which can couple to $D\bar{D^*}+c.c.$ are included in the effective
coupling form factor $g_{\gamma^* D\bar D^*}(s)$. This will lead to
double-counting with the treatment of explicitly including
$\psi(3770)$ and other resonances. Because of this ambiguity we will
only calculate the exclusive open channel contributions in the
following section, and discuss its behavior around the open channel
kinematics.

\subsection{ $\psi^\prime$-$\psi(3770)$ mixing }\label{sec-iic}

The nonvanishing $g_{\psi^\prime D\bar{D}}$ extracted from the cross
section measurement will allow us to investigate the
$\psi^\prime$-$\psi(3770)$ mixing. As a dynamic consequence of the
intermediate $D\bar{D}$ loop (and $D\bar{D^*}+c.c.$ etc), such a
higher order effect can be quantified in $e^+ e^- \to D\bar{D}$.

For two-state $\psi^\prime - \psi(3770)$ mixing, the covariant
propagator can be expressed as~\cite{Achasov:1979xc,Wu:2008hx}:
 \bea G=
\frac{1}{D_{\psi^\prime} D_{\psi(3770)} - |D_{\psi^\prime
\psi(3770)}|^2} \left(
                  \begin{array}{cc}
                    D_{\psi^\prime} & D_{\psi^\prime
\psi(3770)} \\
                    D_{
 \psi^\prime \psi(3770)} & D_{\psi(3770)} \\
                  \end{array}
                \right)
,\eea
 where $D_{\psi^\prime}$ and $D_{\psi(3770)}$ are the
denominators for the propagators of $\psi^\prime$ and $\psi(3770)$:
\be D_{i} = s - m_{i}^2 + i \sqrt{s} ~\Gamma_i(s),\ee
 with the energy-dependent width
 \bea \label{s-dependent-width}
 \Gamma_{i}(s) = \frac{1}{6\pi s} [g_{\psi _i D^+ D^-}^2
|\vec{p}_{D^+}(s)|^3  + g_{\psi _i D^0 \bar D^0}^2
|\vec{p}_{D^0}(s)|^3 + \cdots] + \Gamma_{\psi _i}^{\mbox{non}- D\bar
D},
 \eea
dominated by the $D\bar{D}$ channel near the $\psi(3770)$ mass
region.  $|\vec{p}_{i}(s)| = \sqrt{s - 4 m_i^2}/2 $ is the
three-vector momentum carried by the intermediate $D$ meson at an
energy $\sqrt{s}$.

$D_{\psi^\prime \psi(3770)}$ is the mixing term via the $D\bar D$
meson loop:
 \bea D_{\psi^\prime \psi(3770)} =
\frac{1}{6\pi ~\sqrt{s}} [g_{{\psi^\prime} D^+ D^-} g_{
\psi(3770)D^+ D^-} |\vec{p}_{D^+}(s)|^3 + g_{{\psi^\prime} D^0 \bar
D^0} g_{ \psi(3770)D^0 \bar D^0} |\vec{p}_{D^0}(s)|^3 + \cdots] \ ,
\label{eq-mixing} \eea
 from which we can define the mixing parameter $|\xi|$:
 \be\label{mix-para}
|\xi_i|\equiv \left|\frac{D_{\psi^\prime \psi(3770)}}{D_{i}}\right|
\ . \ee

Several points should be made: (i) In the energy region near the
$\psi(3770)$ mass (below the $D\bar{D^*}+c.c.$ open channel), the
$D\bar{D}$ loop is the dominant contribution to the mixing matrix
element. We restrict the discussion in this region and neglect the
contributions from the $D\bar{D^*}+c.c.$ and other high threshold
loops. But we mention that contributions from the $D\bar{D^*}+c.c.$
loop are rather small when $\sqrt{s}$ is above the threshold. (ii)
In principle, the vertex couplings $g_{\psi^\prime D\bar{D}}$ and
$g_{\psi(3770)D\bar{D}}$ should be $s$-independent in
Eq.~(\ref{mixing}). In our case, the $g_{\psi^\prime D\bar{D}}$ is
extracted by fitting the cross section data, and no obvious
$s$-dependence is needed within the energy region near threshold. We
thus adopt a constant coupling here. Nevertheless, it shows that the
$s$-dependence of the total width $\Gamma_{i}(s)$ is also weak. This
justifies that the mixing parameter extracted above is at an
accuracy of leading order. (iii) In Eq.~(\ref{s-dependent-width})
the non-$D\bar{D}$ channel contributions to the total width are
rather small due to much weaker couplings. As a leading order
estimate we apply the PDG values to fix these two widths and do not
consider their $s$-dependence, i.e. $\Gamma_{\psi
^\prime}^{\mbox{non}- D\bar D} = 317 ~\mbox{KeV}$ and
$\Gamma_{\psi(3770)}^{\mbox{non}- D\bar D} = 4.15
~\mbox{MeV}$~\cite{pdg2008}.

\section{Numerical Results}

Proceeding to the numerical calculation, our fitting strategy is to
minimize the background contributions given that the necessary
resonance amplitudes are included. We thus perform two separate fits
of the data. One is restricted to the near threshold region (Fit-I)
and the other is extended to $\sqrt{s}\simeq 4.3$ GeV covered by the
Belle data~\cite{Pakhlova:2008zza} (Fit-II). This separate treatment
is based on the different experimental situations between BES and
Belle. BES has a relatively detailed scan over the energy region
from threshold to the upper end of the $\psi(3770)$ mass, while the
energy bins in Belle data are much larger and the cross sections
have larger uncertainties. Moreover, the Belle data are extracted
from the exclusive initial state radiation (ISR) production of
$D\bar{D}$ at c.m. energy 10.58 GeV, and are corrected by the ISR.
The advantage of the Belle data is that they cover nearly all
 the energies from threshold up to $\sim 5$ GeV.

\subsection{Fit-I}

Quantum mechanically, the most important contributions to the $e^+
e^-\to D\bar{D}$ cross sections near threshold would come from the
nearby vector charmonia, i.e. $\psi(3770)$ and $\psi^\prime$.
Although $\psi^\prime$ is below the $D\bar{D}$ threshold, its mass
is not located far away. Thus, its off-shell contribution may still
be sizable. Since the $\psi^\prime$ has a well-defined mass position
and total width, we then fix these two quantities in the analysis,
but leave its coupling to $D\bar{D}$ to be fitted by data.

Other fitting parameters include the mass and total width of
$\psi(3770)$, the $\psi(3770)D\bar{D}$ coupling, and the relative
phase angle between the $\psi(3770)$ and $\psi^\prime$ amplitude. By
fitting these parameters to the BES data~\cite{Ablikim:2008zz}, we
actually minimize the contributions from the other resonances and
``bare photon". In Table.~\ref{table1}, the fitting results for
$e^+e^-\to D^0 \bar D^0$ and $ e^+e^-\to D^+ D^-$ are listed
separately. It shows that the values of  $\chi^2$ are compatible in
these two channels. Interestingly, the extracted couplings
$g_{\psi^{\prime} D \bar D}$ and $g_{\psi(3770) D \bar D}$ appear to
have significant differences in the charged and neutral channel. As
a comparison, if we extract $g_{\psi(3770) D \bar D}$ by the PDG
data~\cite{pdg2008}, we obtain $g_{\psi(3770) D^0 \bar{D^0}}=12.43$
and $g_{\psi(3770) D^+ D^-}=12.89$, which are apparently different
from the fitted values. In particular, the BES data suggest a
relatively larger $g_{\psi(3770) D^0 \bar{D^0}}$ coupling than
$g_{\psi(3770) D^+ D^-}$. We note that in both the charged and
neutral channel, the fitted mass and total width of $\psi(3770)$ are
consistent with each other. It is also interesting to note that the
extracted values for $g_{\psi^{\prime} D \bar D}$ are within the
range from other
approaches~\cite{Deandrea:2003pv,Matheus:2002nq,Bracco:2004rx,Lin:1999ad}.

\begin{table}
\caption{ Fitting parameters for  $e^+e^-\to D^0 \bar D^0$ and $
e^+e^-\to D^+ D^-$ near threshold. The experimental data are from
BES~\cite{Ablikim:2008zz}.} \label{table1}
\begin{tabular}{c|c|c}\hline
 & {$e^+e^- \to D^0 \bar D^0$}  & {$e^+e^- \to D^+ D^-$}\\\hline
$g_{\psi^{\prime} D \bar D}$ & $9.05 \pm 2.34$  & $7.72 \pm
1.02$\\\hline $g_{\psi(3770) D \bar D}$ & $13.58 \pm 1.07$  & $10.71
\pm 1.75$\\\hline $m_{\psi(3770} $ (MeV) & $3774.0 \pm 1.3$  &
$3774.0 \pm 1.6$\\\hline $\Gamma_{\psi(3770)}$(MeV) & $28.4 \pm 2.9$
& $29.6 \pm 2.9$\\\hline $\phi$& $228.1^\circ \pm 18.6^\circ$ &
$190.6^\circ \pm 21.1^\circ$
\\\hline $\chi^2/$d.o.f& 12.28/9 & 13.32/9\\\hline
\end{tabular}
\end{table}

\begin{figure}[ht]
\begin{center}
\begin{tabular}{ccccc}
\scalebox{0.6}{\includegraphics{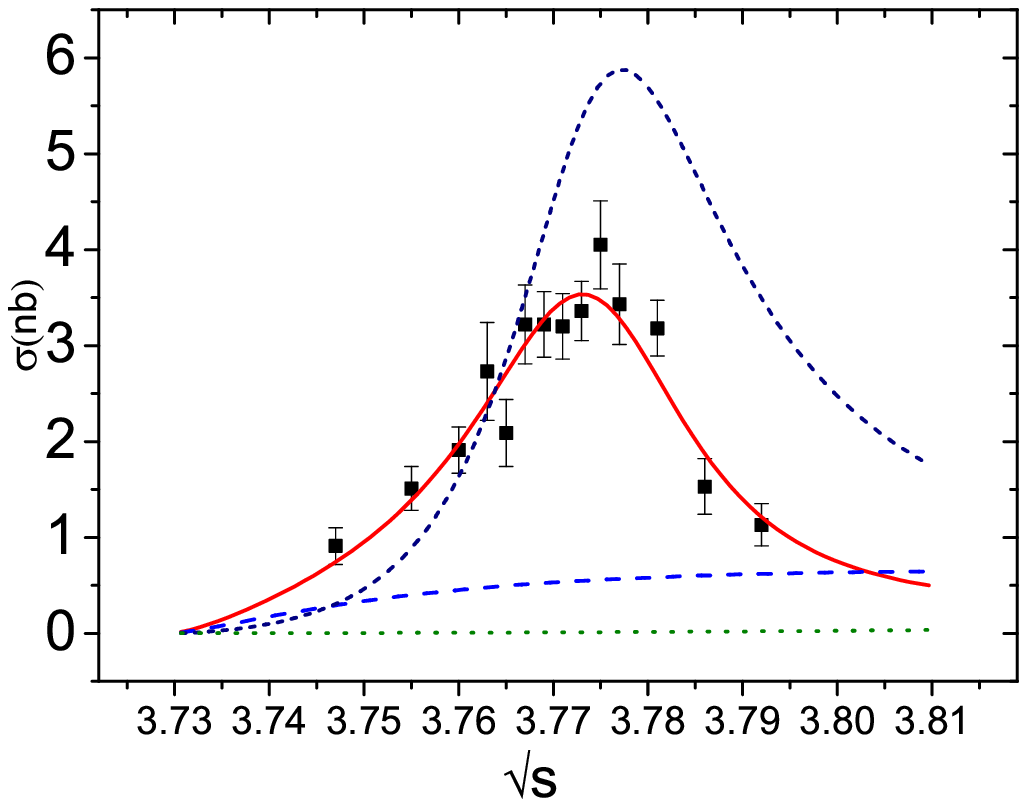}}&&\scalebox{0.6}{\includegraphics{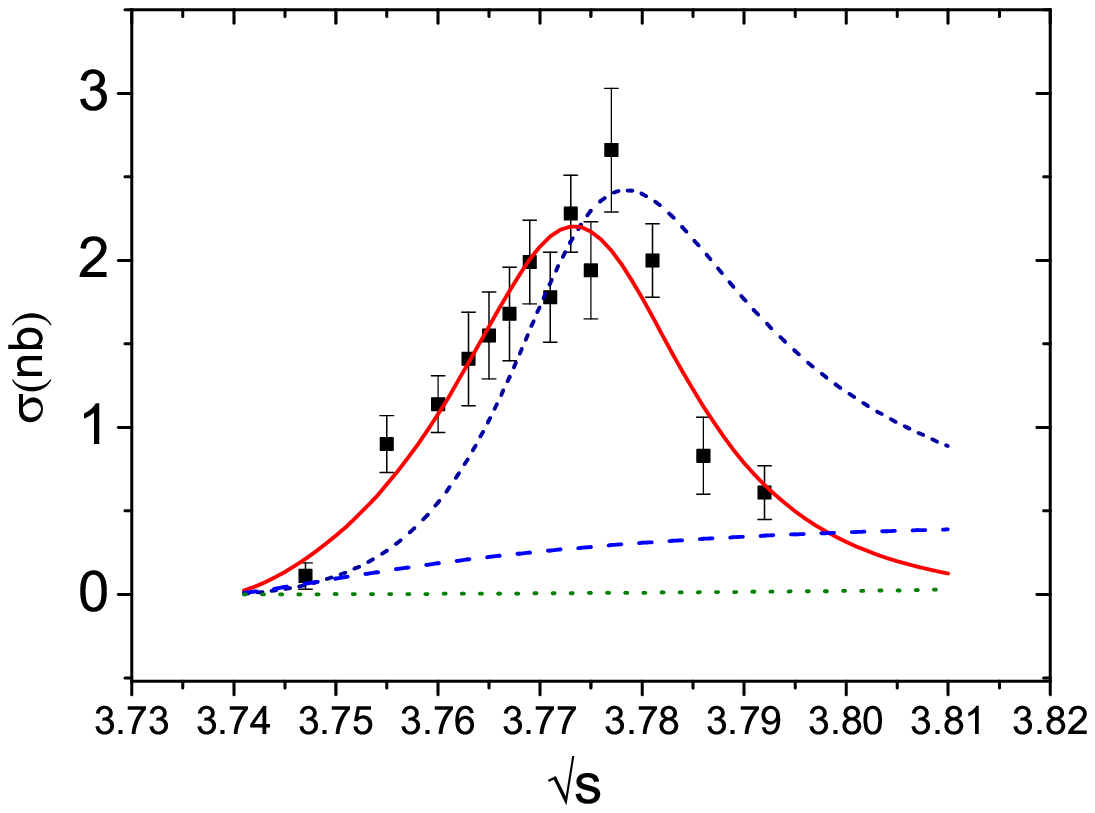}}\\
(a)&&(b)\\
\end{tabular}
\end{center}
\caption{The cross sections for $e^+e^- \to D^0\bar D^0$ (left
panel) and $e^+e^- \to D^+ D^-$ (right panel) fitted by the
$\psi^\prime$ and $\psi(3770)$ interferences (solid line). The
short-dashed line stands for the exclusive $\psi(3770)$ cross
section, while the dashed line denotes the one for $\psi^\prime$.
The dotted line represents the exclusive contribution from
$\psi(4040)$. The data are from BES~\cite{Ablikim:2008zz}.}
\label{resp}
\end{figure}

In Fig.~\ref{resp}, the fitting results for both channels are
presented. The short-dashed curves are exclusive $\psi(3770)$ cross
sections while the long-dashed curves are exclusive $\psi^\prime$.
The solid curves are given by the interfering amplitudes between
$\psi(3770)$ and $\psi^\prime$ with the relative phase angle
$\phi=228.1^\circ$ and $190.6^\circ$ for the neutral and charged
channel, respectively. Apart from the coupling differences, the
differences between the phase angles also significantly contribute
to the interferences.

As an estimate of higher resonance effects, we take the PDG average
of the total width and the upper limit of the branching ratio
${\mbox{Br}}(\psi(4040)\to D \bar{D}) < 0.2\%$~\cite{pdg2008} to
extract $g_{\psi(4040) D\bar D} \approx 0.34 $. The $\psi(4040)$
decay constant can also be extracted from the experimental data by
Eq.~(\ref{eq2}), which gives $e/f_{\psi(4040)} = 9.35\times
10^{-3}$. In Fig.~\ref{resp} the exclusive cross sections from
$\psi(4040)$ are presented as the dotted curves, which turn out to
be negligible near threshold. This justifies the treatment that the
cross sections from threshold to the upper end of the $\psi(3770)$
mass are dominated by $\psi(3770)$ and $\psi^\prime$ interferences,
and the background contributions (here, they are referred to the
``bare photon" and higher resonances) are minimized.

\begin{figure}[ht]
\scalebox{0.7}{\includegraphics{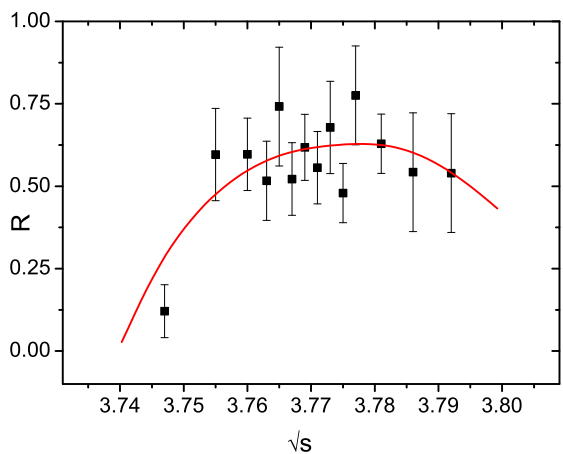}}
\caption{ The cross section ratio between $e^+ e^- \to D^+D^-$ and $
e^+ e^- \to D^0\bar{D^0}$ versus c.m. energy $\sqrt{s}$. The data
are from BES~\cite{Ablikim:2008zz}.} \label{ratio}
\end{figure}

We also calculate the energy-dependence of the cross section ratio
between $e^+e^- \to D^+ D^-$ and $e^+e^- \to D^0 \bar D^0$ in
Fig.~\ref{ratio} to compare with the BES data. This shows an overall
agreement, except that we should note that the ratio may start to
deviate from reality at higher energies since other resonances and
mechanisms would start to have stronger interferences. Since the
cross sections appear to be sensitive to tiny discrepancies between
the charged and neutral channel, a more precise measurement of this
quantity would be useful for understanding the underlying
mechanisms.

\subsection{Fit-II}

In this subsection, we fit the experimental data from
Belle~\cite{Pakhlova:2008zza} by including higher resonances. Apart
from $\psi(4040)$ and $\psi(4160)$ from PDG~\cite{pdg2008}, we also
include a new state $X(3900)$ in the fit. As shown by the Belle data
for $e^+ e^- \to D^0 \bar D^0$~\cite{Pakhlova:2008zza}, an
enhancement is observed at about 3.9 GeV which could be signals for
a charmonium state. Although we fit the data using a Breit-Wigner
here, we shall argue later that this enhancement may be due to the
$D\bar{D^*}+c.c.$ open channel effects.

The strategy is similar to Fit-I. Namely, we fix the mass and total
width of the $\psi^\prime$ and then leave $g_{\psi^\prime
D^0\bar{D^0}}$ to be fitted by data. The other fitted quantities
include the resonance parameters for $\psi(3770)$, $X(3900)$,
$\psi(4040)$, $\psi(4160)$, and their relative phase angles to the
$\psi^\prime$ amplitude. In Table.~\ref{table2}, the fitting results
are listed and can be compared with Table.~\ref{table1}.

This shows that the coupling $g_{\psi^\prime D^0\bar{D^0}}$ appears
to be stable in this fit, while parameters for $\psi(3770)$ exhibit
strong sensitivities to the data. In particular, the total width
becomes rather small, even smaller than the PDG
average~\cite{pdg2008}. The phase angle also changes, and the
overall effects are that the cross sections at the $\psi(3770)$ mass
are increased. Such a dramatic change is likely due to the
significant discrepancies between the BES and Belle data. Again, a
detailed scan over the threshold region is strongly required. The
fitted total widths for $X(3900)$ and $\psi(4040)$ turn out to have
large uncertainties.

\begin{table}
\caption{ Fitting parameters from Fit-II. The results are obtained
by fitting the Belle data~\cite{Pakhlova:2008zza} for $e^+e^-\to D^0
\bar D^0$. We separately list the other two fitted quantities here:
$g_{\psi^\prime D^0 \bar D^0}= 9.05 \pm 0.37$, $g_{\psi(3770) D^0
\bar D^0} = 13.13 \pm 0.65$. The reduced $\chi^2$ is
$\chi^2/\mbox{d.o.f}=17.45/19$. } \label{table2}
\begin{tabular}{c|c|c|c|c}
\hline  & $M$ (GeV) & $\Gamma$ (MeV)
 &  $ g_{VDD}/f_V  (\times 10^{-2})$  & $\phi$ \\
\hline X(3900) & 3.894 $\pm$ 0.011& 89.8 $\pm$ 12.6
 & 6.76 $\pm$ 0.89& $104.36^\circ \pm 7.90^\circ$\\
\hline $\psi(3770)$ & 3.7724 $\pm$ 0.002 & 25.4 $\pm$ 1.4 & 23.2
$\pm$  1.1 & $198.85^\circ \pm 4.19^\circ$
\\\hline
$\psi(4040)$ & 4.0812 $\pm$ 0.008 & 96.2 $\pm$ 11.4
 & 3.48 $\pm$ 0.34& $101.16^\circ \pm 11.742^\circ$\\\hline
\end{tabular}
\end{table}

\begin{figure}[ht]
\scalebox{0.7}{\includegraphics{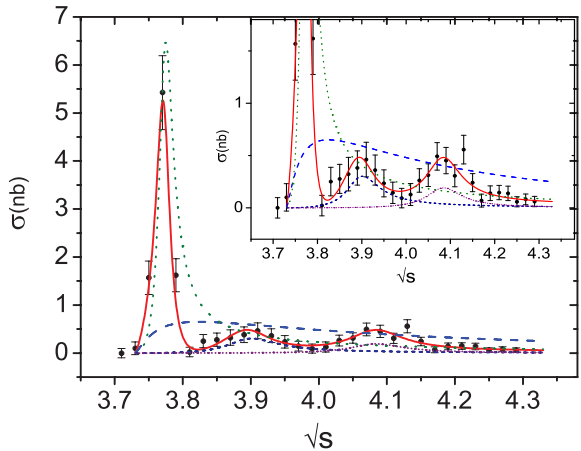}}
\caption{The Belle cross section for $e^+e^- \to D^0 \bar
D^0$~\cite{Pakhlova:2008zza} fitted by Fit-II. The solid line
represents the overall results, the long-dashed line denotes the
exclusive contribution from $\psi^\prime$, and the long-dotted,
short-dashed and short-dotted lines denote the exclusive cross
sections from $\psi(3770)$, $X(3900)$ and $\psi(4040)$,
respectively. } \label{d0d0barbelle}
\end{figure}

We plot the fitting results in Fig.~\ref{d0d0barbelle}, which appear
to be in good agreement with the Belle data. In comparison with
Fit-I, the importance of $\psi^\prime$ is consistently highlighted.
Exclusive cross sections from other resonances are also presented in
Fig.~\ref{d0d0barbelle}. We note that the smooth-out behavior at the
high energy end is favored by the minimization, and it leads to a
negligibly small contributions from $\psi(4160)$. Note that the data
still have rather large uncertainties. The contributions from
$\psi(4160)$ should be restudie with high-quality data, and we will
not discuss them much here.

\subsection{The $D\bar{D^*}+c.c.$ open channel effects}

The enhancement at 3.9 GeV from the Belle data, if is due to
resonance excitation, may cause some confusion about the potential
model predictions~\cite{Eichten:2007qx} for the charmonium spectrum
since one would not expect a $J^{PC}=1^{--}$ state in this mass
region. Taking into account the success of the potential quark
model, an alternative explanation for this structure would be due to
the $D\bar{D^*}+c.c.$ open channel effects as illustrated by
Fig.~\ref{open-channel}.

\begin{figure}[ht]
\scalebox{0.7}{\includegraphics{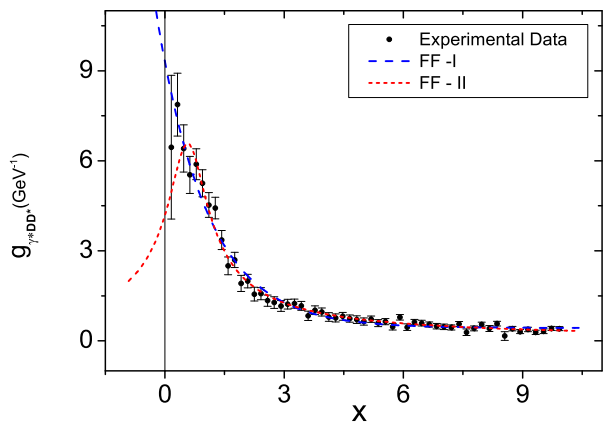}}
\caption{The $x$-dependence of $g_{\gamma^* D\bar D^*}$ extracted
from the cross sections for $e^+ e^- \to D^+ D^{*-} + c.c.$
~\cite{Abe:2006fj} with $x \equiv  s - (m_D + m_{D^*})^2$. The
dashed and dotted line are given by the FF-I and FF-II schemes,
respectively. The vertical line at $x=0$ labels the $D\bar{D^*}$
threshold. }\label{coupling}
\end{figure}

With the help of Eq.~(\ref{crosssection1}) and experimental data for
$e^+ e^- \to D^+ D^{*-} + c.c.$ from Belle~\cite{Abe:2006fj}, the
effective coupling $g_{\gamma^* D \bar D^* }(s)$ can be extracted.
The corresponding datum points are plotted in Fig.~\ref{coupling} in
terms of $x \ (\equiv s-(m_D+m_{D^*})^2)$.

We employ two methods to extract the effective coupling form factor:

(i) Form factor one (FF-I)

In the FF-I scheme, we fit the data in Fig.~\ref{coupling} with an
exponential function:
\begin{equation}
g_{\gamma^* D \bar D^* }(s) = g_1 \exp{[-(s-(m_D+m_{D^*})^2)/t_1]} +
g_0 \ ,
\end{equation}
where $x=0$ corresponds to the $D\bar{D^*}+c.c.$ threshold, and
$g_1$, $t_1$, and $g_0$ are fitting parameters. The above function
can perfectly describe the energy dependence of the effective
coupling. By simply extrapolating the exponential to the $D\bar{D}$
threshold (dashed line in Fig.~\ref{coupling}), we then apply this
form factor to calculate the open channel cross sections.

(ii) Form factor two (FF-II)

In the FF-II scheme, we fit the data in Fig.~\ref{coupling} with a
single resonance:
\begin{eqnarray}
g_{\gamma^* D \bar D^*}(s) = \left|\frac{b_0}{s - m_X ^2 + i m_X
\Gamma _X} + b_1\right| \ ,
\end{eqnarray}
with a background term $b_1$. The parameter $b_0$ can be regarded as
the product of the $\gamma^* X$ coupling and the $XD\bar{D^*}$
coupling. This parametrization agrees with the data at higher
energies, but drops at the threshold region.

\begin{table}
\caption{ Parameters fitted with FF-I and FF-II for the $\gamma^*
D\bar{D^*}$ effective coupling form factor. The data are from
Belle~\cite{Abe:2006fj}.} \label{ff-para}
\begin{tabular}{|c|c|c|c|}\hline
\multicolumn{2}{|c|}{FF-I}
          & \multicolumn{2}{c|}{FF-II}\\\hline
\multicolumn{2}{|c|}{$~~~~g_{\gamma^* D \bar D^*}=g_1
\mbox{exp}[-(s-(m_D + m_{D^*})^2)/t_1] + g_0~~~~$} &
\multicolumn{2}{c|}{ $~~~~g_{\gamma^* D \bar D^*} = |\frac{b_0}{s -
m_X ^2 + i m_X \Gamma _X} + b_1|~~~~$}\\\hline $~~~~g_1~~~~$ & $8.86
\pm 0.59$ & $b_0$ & $3.08 \pm 0.31$
\\\hline
$t_1$ & $1.28 \pm 0.06$ & $m_X ~(\mbox{GeV})$ & $3.943 \pm 0.014$
\\\hline
$g_0$ & $0.43 \pm 0.02$ & $\Gamma_X ~(\mbox{MeV})$ & $119.0 \pm
10.0$
\\\hline
\multicolumn{2}{|c|}{} & $b_1$ & $0.016 \pm 0.045$
\\\hline
$~~~~\chi^2$/d.o.f ~~~~& 68.86/53  & $~~~~\chi^2$/d.o.f ~~~~  &
47.67/52
\\\hline
\end{tabular}
\end{table}

The above effective couplings should be taken with caution. In
principle, the parametrization of the FF-I scheme contains all the
contributions from the vector resonances, while the FF-II scheme
contains only one resonance plus a background. Comparing these two
form factors with each other would suggest the dominance of one
single resonance around 3.94 GeV, which seems to support the need
for $X(3900)$. However, it should be recognized that the
Breit-Wigner structure fitted by the form factor is largely due to
the collective contributions from nearby resonances such as
$\psi(4040)$. The parametrization somehow tilts the resonance signal
due to the unphysical background term. In fact, both
Belle~\cite{Abe:2006fj} and BABAR~\cite{:2009xs} observe the
dominant $\psi(4040)$ in $e^+ e^-\to D\bar{D^*}+c.c.$, but there are
no signs for the $X(3900)$ there. This could, in fact, support the
fact that the enhancement at 3.9 GeV in $e^+ e^-\to D\bar{D}$ is due
to non-resonant mechanisms. For our purpose of extracting the
effective $\gamma^* D\bar{D^*}(s)$ form factor, these
parametrizations should be acceptable.

\begin{figure}[ht]
\scalebox{0.7}{\includegraphics{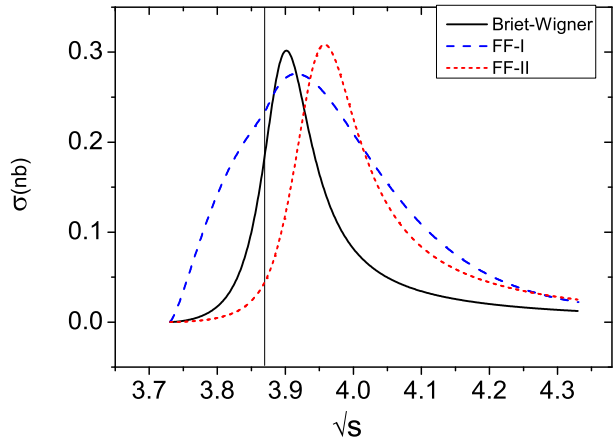}}
\caption{The comparison of cross sections from different mechanisms
around 3.9 GeV. The solid line is given by a Breit-Wigner resonance
$X(3900)$ in Tab.~\ref{table2}, while the dashed and dotted line are
given by the FF-I and FF-II schemes, respectively. The vertical line
labels the $D\bar{D^*}$ threshold. $\alpha = 2.5$ has been taken.}
\label{rescattering}
\end{figure}

With the form factor coupling $g_{\gamma^* D \bar D^* }(s)$, we can
then evaluate the $D\bar{D^*}+c.c.$ open channel effects in $e^+
e^-\to D\bar{D}$. We plot the cross sections from the $D \bar D^*$
open channel in Fig.~\ref{rescattering}. As a comparison, we also
include the Breit-Wigner form factor (solid line) for $X(3900)$.
Taking the cut-off parameter $\alpha = 2.5$, the dashed and dotted
line are given by those two form factors, FF-I and FF-II,
respectively. This shows that FF-I can lead to some enhancement to
the cross sections below $\sim 3.9$ GeV, while the cross sections
from these two form factors are similar to each other above 3.9 GeV.
This shows that the behavior of the solid line is similar to FF-II
except that the fitted mass and width are slightly different. The
reason again should be that the fitted form factor FF-II contain all
contributions from several nearby states. Thus, the fitted
Breit-Wigner parameters contain unphysical information.

It is interesting to see that the cross section peaks from these
different treatments are located in a similar place. In this sense,
we argue that the present data cannot eliminate the possibility that
the cross section enhancement at around 3.9 GeV could be from the
$D\bar{D^*}$ open channel contributions.

\subsection{$\psi^\prime$-$\psi(3770)$ mixing results}

\begin{figure}[ht]
\scalebox{0.7}{\includegraphics{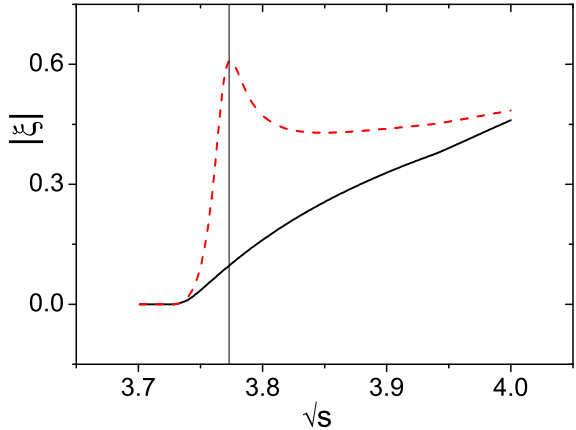}}
\caption{The energy-dependence of the mixing parameter. The solid
line is for $|\xi_{\psi^\prime}(s)|$ while the dashed line for
$|\xi_{\psi(3770)}(s)|$. The vertical line labels the $\psi(3770)$
mass position. }\label{mixing}
\end{figure}

Following the discussion of Sec.~\ref{sec-iic}, we now investigate
the energy dependence of the mixing parameter $|\xi_i|$ defined in
Eq.~(\ref{mix-para}). As mentioned earlier, the $s$-dependence of
the total widths is weak for both $\psi^\prime$ and $\psi(3770)$
near threshold; we thus keep the widths as constants. The
$g_{\psi(3770) D\bar D}$ coupling appear to be different in the
charged and neutral channels. Here, for the purpose of investigating
the energy-dependence of the mixing parameter, such a difference can
be neglected. We then take $g_{\psi^\prime D\bar D} = 9.05$ (the
fitting results) and $g_{\psi(3770) D\bar D} = 12.66$ (the average
obtained from Ref.~\cite{pdg2008}) to extract the mixing parameter
$|\xi_i|$. The results are presented in Fig.~\ref{mixing}.

The solid line denotes the mixing amplitude
$|\xi_{\psi^\prime}(s)|\equiv |D_{\psi^\prime
\psi(3770)}/D_{\psi^\prime}|$, while the dashed line is for
$|\xi_{\psi(3770)}(s)|\equiv |D_{\psi^\prime
\psi(3770)}/D_{\psi(3770)}|$. There are clear physical meanings for
these two quantities. At a given energy $\sqrt{s}$, the value of
$|\xi_{\psi^\prime}(s)|$ denotes the coupling strength of the
$\psi^\prime$ component inside an initial $\psi(3770)$ state
(propagator), while $|\xi_{\psi(3770)}(s)|$ denotes the strength of
the $\psi(3770)$ inside an initial $\psi^\prime$ state.

The mixing parameter $|\xi_{\psi^\prime}(s)|$ at $\sqrt{s}=3.773$
GeV can be related to the $\psi(2S)$-$\psi(1D)$ state mixing angle
via $|\xi_{\psi^\prime}(s)|\simeq
|\sin\theta_{\psi^\prime}|$~\cite{Rosner:2004wy} for which we find
$\theta_{\psi^\prime} \simeq 5.4^\circ$. This value agrees with the
one extracted in Ref.~\cite{Zhang:2009kr}, although it is only about
half of the result estimated by Ref.~\cite{Rosner:2004wy}. It is
worth emphasizing that this quantity strongly depends on the
coupling strength of the $g_{\psi^\prime D\bar{D}}$ coupling which
is a consequence of the lineshape measurement of the cross sections
near threshold. In this sense, the cross section for $e^+ e^-\to
D\bar{D}$ directly provides a constraint on the mixing angle.

\section{Summary and discussion}

In this work, we investigate the reaction mechanisms for $e^+ e^-\to
D\bar{D}$ from threshold to $\sqrt{s}\simeq 4.3$ GeV in an effective
Lagrangian approach. We find that the cross section lineshape near
threshold is very sensitive to the presence of $\psi^\prime$. By
fitting the cross sections from BES~\cite{Ablikim:2008zzc} and
Belle~\cite{Pakhlova:2008zza}, we succeed in extracting resonance
parameters for $\psi^\prime$, $\psi(3770)$, $X(3900)$, $\psi(4040)$.
The shows that the coupling $g_{\psi^\prime D\bar D}$ is consistent
with those values given by other analyses. We show that it is
important to have a reliable determination of this quantity, not
only to understand the lineshape of the cross section at the mass of
$\psi(3770)$, but also to provide a probe for the
$\psi^\prime$-$\psi(3770)$ state mixing.

The analysis also suggests a significant difference between
$g_{\psi(3770)D^0\bar{D^0}}$ and $g_{\psi(3770)D^+ D^-}$, which is
quite different from the PDG averaged values~\cite{pdg2008}. In
order to further clarify this, a precise measurement of both the
$D^0\bar{D^0}$ and $D^+ D^-$ cross sections is strongly recommended,
and BES-III would have great advantages regarding this
issue~\cite{ronggang}.

We also study the $D\bar{D^*}$ open channel effects on the $e^+
e^-\to D\bar{D}$ cross sections. We find that we cannot eliminate
the fact that the enhancement at 3.9 GeV in the Belle data is due to
the $D\bar{D^*}$ open channel contributions. Further investigation
of this mechanism with more accurate data would be needed.

\section*{Acknowledgement}

Useful discussions with Gang Rong and X.L. Wang are acknowledged. We
thank Galina Pakhlova for useful comments on this work. This work is
supported, in part, by the National Natural Science Foundation of
China (Grants No. 10675131 and 10491306), Chinese Academy of
Sciences (KJCX3-SYW-N2), and Ministry of Science and Technology of
China (2009CB825200).

\end{document}